# A TWO-STAGE INJECTION-LOCKED MAGNETRON FOR ACCELERATORS WITH SUPERCONDUCTING CAVITIES*

Grigory Kazakevich#, Rolland Johnson, Gene Flanagan, Frank Marhauser, Mike Neubauer,
Muons, Inc., Batavia, 60510 IL, USA
Vyacheslav Yakovlev, Brian Chase, Sergey Nagaitsev, Ralph Pasquinelli, Nikolay Solyak,
Vitali Tupikov, Daniel Wolff, Fermilab, Batavia, 60510 IL, USA

*Abstract*

A concept for a two-stage injection-locked CW magnetron intended to drive Superconducting Cavities (SC) for intensity-frontier accelerators has been proposed. The concept considers two magnetrons in which the output power differs by 15-20 dB and the lower power magnetron being frequency-locked from an external source locks the higher power magnetron. The injection-locked two-stage CW magnetron can be used as an RF power source for Fermilab's Project-X to feed separately each of the 1.3 GHz SC of the 8 GeV pulsed linac. We expect output/locking power ratio of about 30-40 dB assuming operation in a pulsed mode with pulse duration of ≈ 8 ms and repetition rate of 10 Hz. The experimental setup of a two-stage magnetron utilising CW, S-band, 1 kW tubes operating at pulse duration of 1-10 ms, and the obtained results are presented and discussed in this paper.

## INTRODUCTION

Intensity frontier high-energy accelerators of proton and ion beams are crucial for Accelerator Driven Systems, ADS, (intended to drive sub-critical reactors, ADSR) and research projects including the Project X, which is an accelerator facility under development at Fermilab [1], developed to support multiple physics programs at the intensity frontier. Additionally, Project X could be used to advance the concepts for an ADSR. Subsequent stages of Project X could include an 8 GeV pulsed linac for future muon and/or neutrino facilities.

The pulsed linac will utilize the standard 1.3 GHz ILC-type SC [2], with an accelerating field of 25 MV/m. The cavity filling time and the full RF pulse duration are chosen to be about 3 ms and 8 ms respectively. The total number of cavities is 200. With a beam current of ~1 mA the power required to feed each cavity is ~ 40 kW.

Several options for driving the SC are being considered. One option is based on a single high power klystron feeding a number of SC (16 cavities in 2 cryomodules). However the scheme does not allow independent tuning of the phase and accelerating voltage in the cavities which is highly important for weakly-relativistic particles. Another option is based on individual feeding of each cavity from a separate IOT or klystron transmitter. This option looks technically plausible, but the number of transmitters significantly increases project cost. Thus, a potentially less expensive RF source to excite each cavity independently is an important option to be considered for Project X and other SC accelerators. The concept of such an RF source and its modelling are discussed in the paper.

*Work is supported by the US DOE grant DE-SC0006261
#grigory@muonsinc.com; gkazakevitch@yahoo.com

## AN INJECTION-LOCKED MAGNETRON AS AN RF SOURCE FOR SC ACCELERATORS

A general requirement for the state of the art intensity frontier SC accelerators of weakly-relativistic particles is high phase stability of the RF source feeding the SC. The allowable phase instability of ~1 degree, [3], results from the beam transverse and longitudinal emittance simulations to avoid beam loss. Also, it is necessary that the accelerating voltage be stable to within 1%, [ibid]. To provide phase stabilization of the RF sources it is necessary to use feedback of a Low Level RF (LLRF) system. Power control, necessary for operation with SC, can be done as proposed in [4].

In the context of magnetron RF sources, intended to drive the SC, this task is formally similar to consideration of a frequency-locked magnetron controlled by a slowly varying phase/frequency. Such a task has been successfully solved theoretically using a numerical simulation of transient processes in the frequency-locked magnetron considering it as a forced oscillator [5-7]. Experimental validation of the simulation was performed with a 2.5 MW S-Band pulsed magnetron [ibid]. Figure 1 shows excellent agreement between simulation and measurements [6, 7] for the locking power of -18 dB.

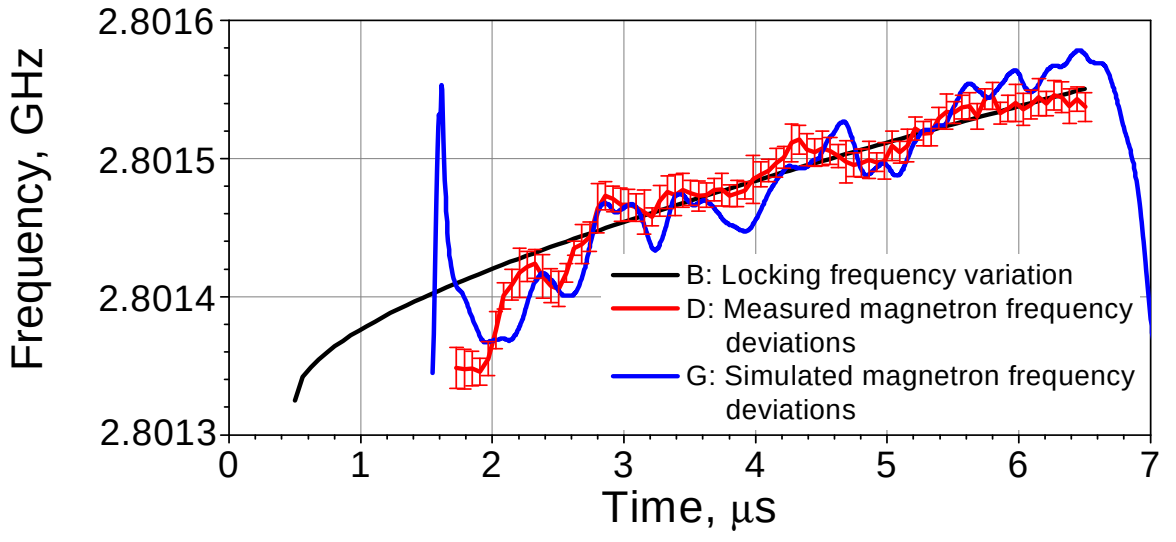

Figure 1: Variation of the simulated (G), and measured magnetron frequency (D), vs. the locking frequency (B).

Analysis of results obtained in [5-7] was subsequently used to compute the phase deviations of the 2.5 MW magnetron locked by the slowly varying frequency at power of -18 dB. Figure 2 shows simulated phase deviations of the frequency-locked magnetron relatively to the locking frequency, curve E, and the measured deviations, curve G, which are in agreement as well. The intrapulse phase deviations are ~1 degree; such stability is adequate for the Project X pulsed linac requirements.

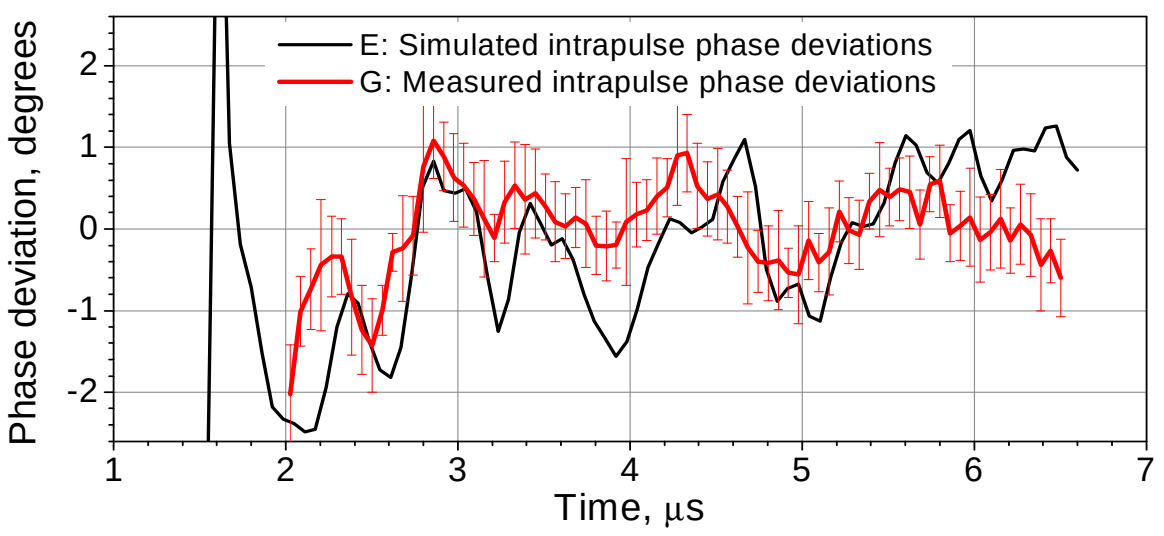

Figure 2: Simulated, E, and measured, G, intrapulse phase deviations of the 2.5 MW frequency-locked magnetron.

These results, and the experimental results obtained with a commercial injection-locked CW magnetron [8, 9], motivated our concept of the 2-stage (cascade) magnetron for driving the SC of intensity frontier accelerators.

## EXPERIMENTS WITH A SINGLE-STAGE FREQUENCY-LOCKED MAGNETRON

The experiments were performed with commercial 2.45 GHz magnetrons with output power up to 1 kW operated in long pulse duration mode. The magnetron with corresponding RF components was assembled as a self-contained module, Figure 3.

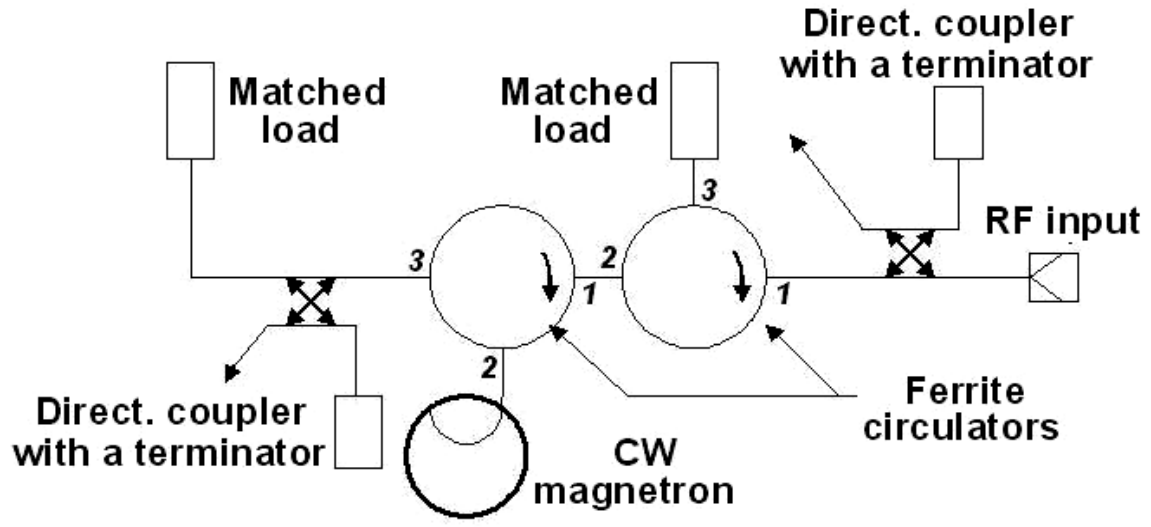

Figure 3: A single CW magnetron experimental module operating as a pre-excited frequency-locked oscillator.

The magnetron module was fed by a pulsed modulator using partial discharge of a storage capacitor. The modulator voltage slope was less than 0.9% when the modulator was loaded concurrently by 2 magnetrons and the pulse duration was 10 ms. The magnetron was pre-excited by a precise CW oscillator with a TWT amplifier. First measurements were done using down-conversion with an intermediate frequency (IF) = 10 kHz and a magnetron pulse duration of 2.5 ms. A vector network analyzer (HP 8753ES) synchronized with the type N5181A CW oscillator was used as the heterodyne. Phase deviations were measured and found to be $\delta\varphi$ (rms) ≈1.2 deg. during the main part of the magnetron pulse, demonstrating good phase stability of the frequency-locked magnetron in the pulsed regime.

Further measurements were performed with an interferometer, Figure 4, for pulse duration of ≈8 ms. The scheme reduced the phase noise caused by the vector network analyzer in the down-conversion measurements. The measurements were performed at a locking power and a magnetron output power of: $P_{Lock}$ ≈7.3 W and $P_{Out}$ ≈ 580 W respectively (or $P_{Out}/P_{Lock}$ ≈19 dB).

Measured phase deviations are plotted in Figure 5, a-d.

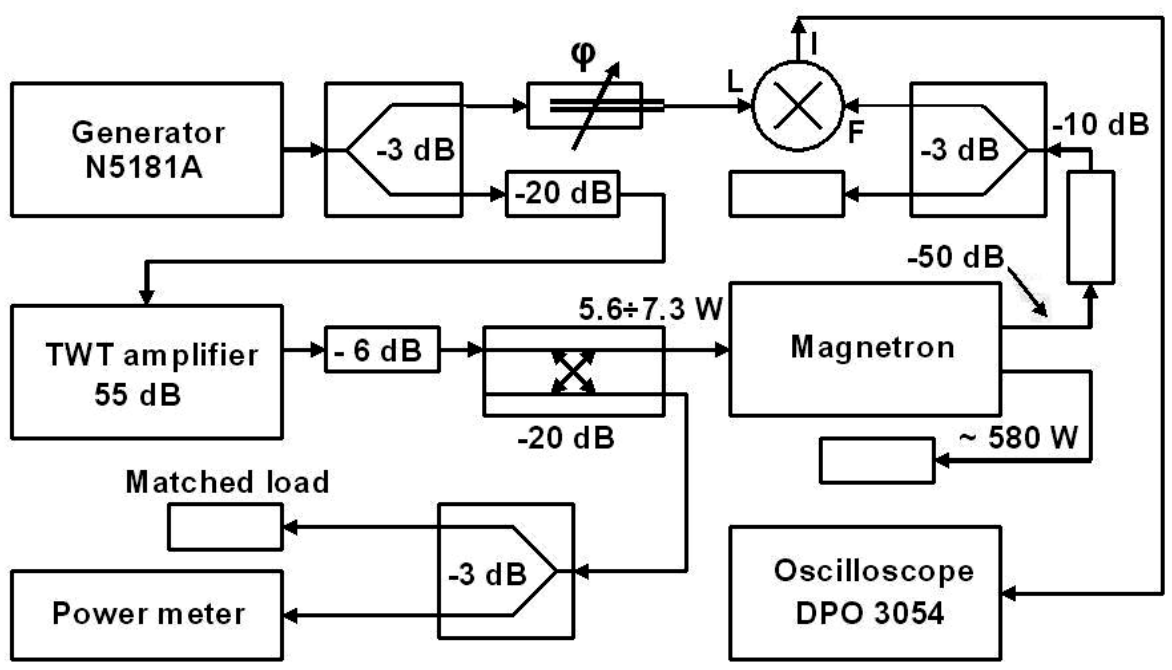

Figure 4: Scheme with an interferometer to measure phase deviations of the frequency-locked magnetron.

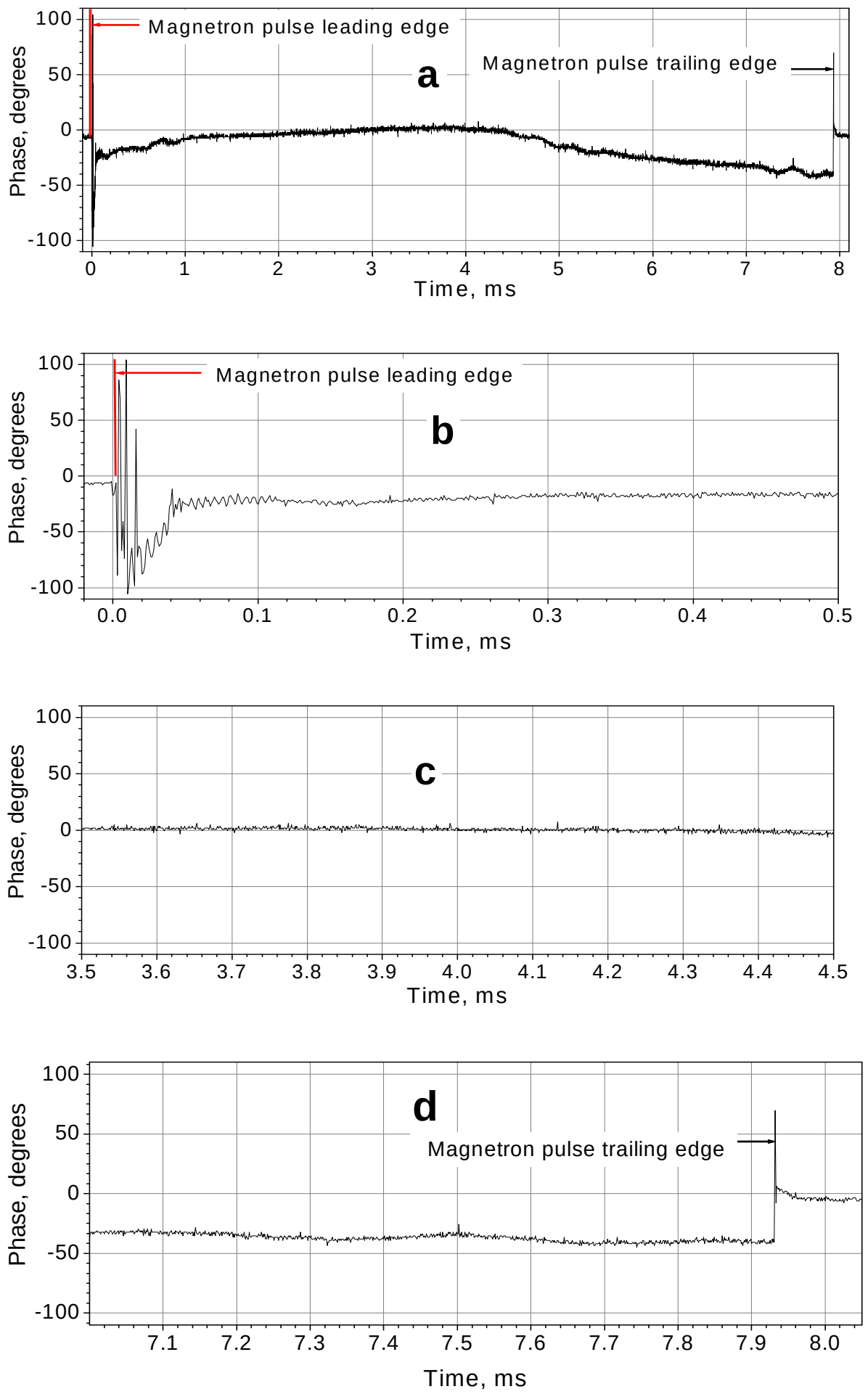

Figure 5, a-d: Phase deviations of the frequency-locked CW magnetron operating with pulse duration of 8 ms.

The first plot shows phase deviations during the full 8 ms pulse. Figure 5b shows phase deviations in beginning of the pulse, it can be seen that the time-to-lock is ≤50 µs. The subsequent plots show phase deviations in middle and last parts of the 8 ms pulse. All the plots show instantaneous phase deviations of less than a few degrees in main part of the pulse (t ≥120 µs) for the pre-excited frequency-locked magnetron operating in pulse mode. Slow drift of the phase during the long pulse is explained by two competing processes: an increase of the magnetron current because of overheating of the cathode caused by bombardment of returning electrons, and a decrease of the magnetron current associated with a drop in the magnetron voltage because of a discharge of the storage capacity. The later process dominates in the second half of the pulse. The obtained phase deviations are acceptable for a phase control with a LLRF system.

## EXPERIMENTS WITH A TWO-STAGE FREQUENCY-LOCKED MAGNETRON

A model of the two-stage (cascade) magnetron, proposed to drive the SC of the Project X pulsed linac and other intensity frontier accelerators, was built with two similar CW 2.45 GHz magnetrons with output power up to 1 kW operating in pulsed regime. The magnetrons were chosen to be locked at the same frequency. Both magnetron modules were fed by a single modulator, previously used for the single magnetron measurements. The magnetron with the lower nominal anode voltage was fed through a divider. The modulator pulse duration was chosen to be 5 ms to avoid the aforementioned noticeable drop of the magnetron current. The first magnetron was frequency-locked by a CW signal; the second one was excited by the output signal of the first one through an attenuator lowering the locking power for the second magnetron. The scheme for the phase measurements with the 2-stage (cascade) magnetron is shown in Figure 6.

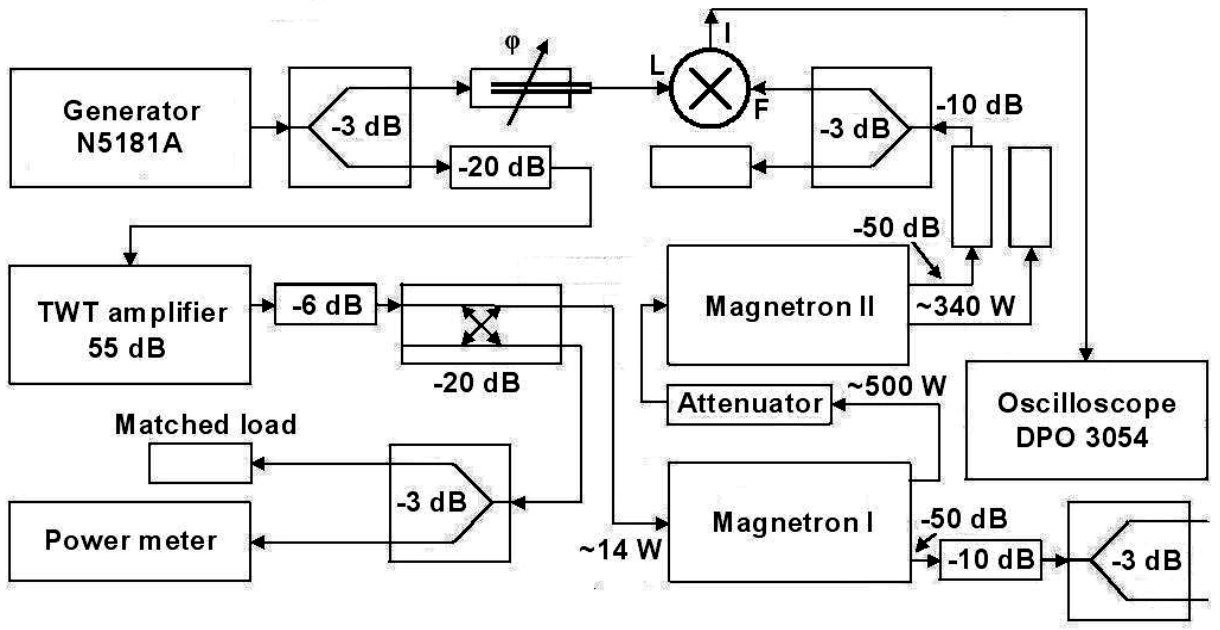

Figure 6: Scheme for measurements of the phase stability of the 2-stage CW magnetron in pulsed regime.

The measurements were performed at attenuation of -20 dB and -13 dB corresponding to the ratio of $P_{Out-II}/P_{Lock-I} \approx$ 33.5 dB and 26.5 dB, respectively. The phase variation for the 5 ms pulse duration are shown in Figure 7, a-d.

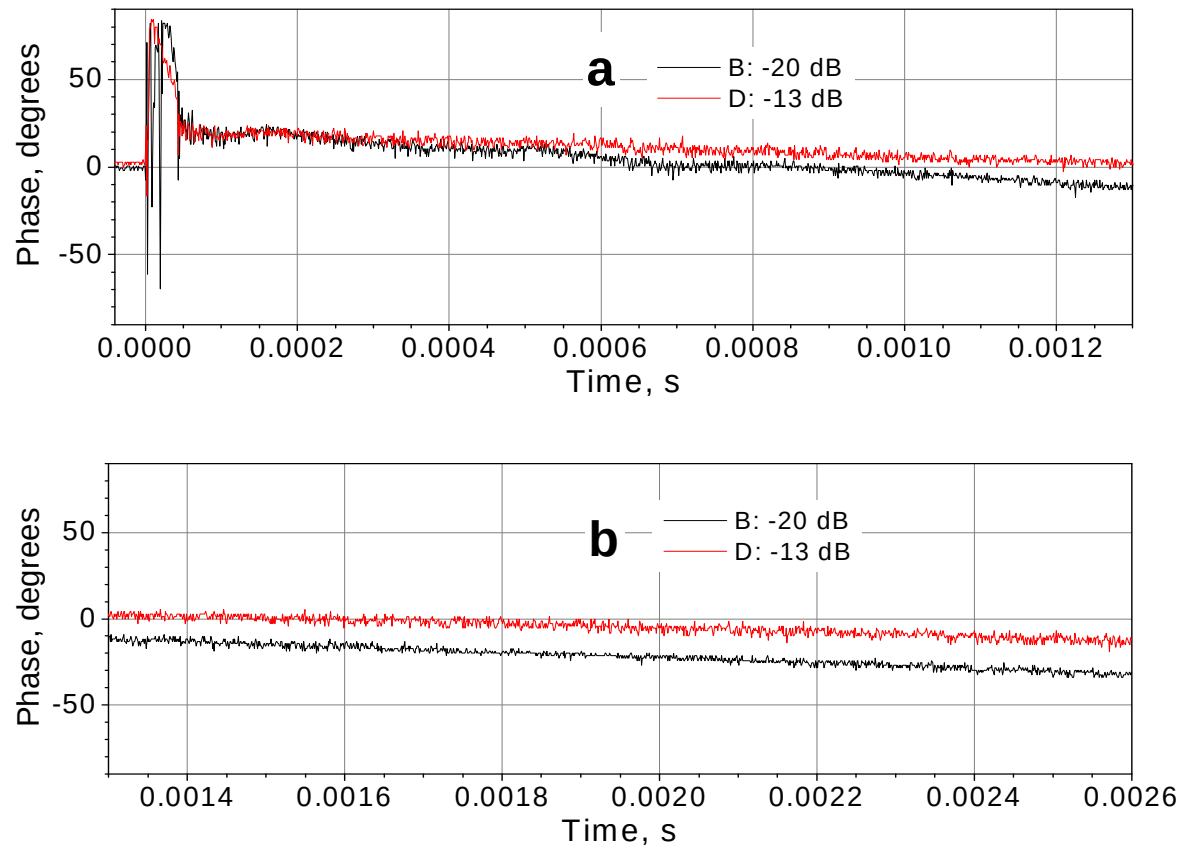

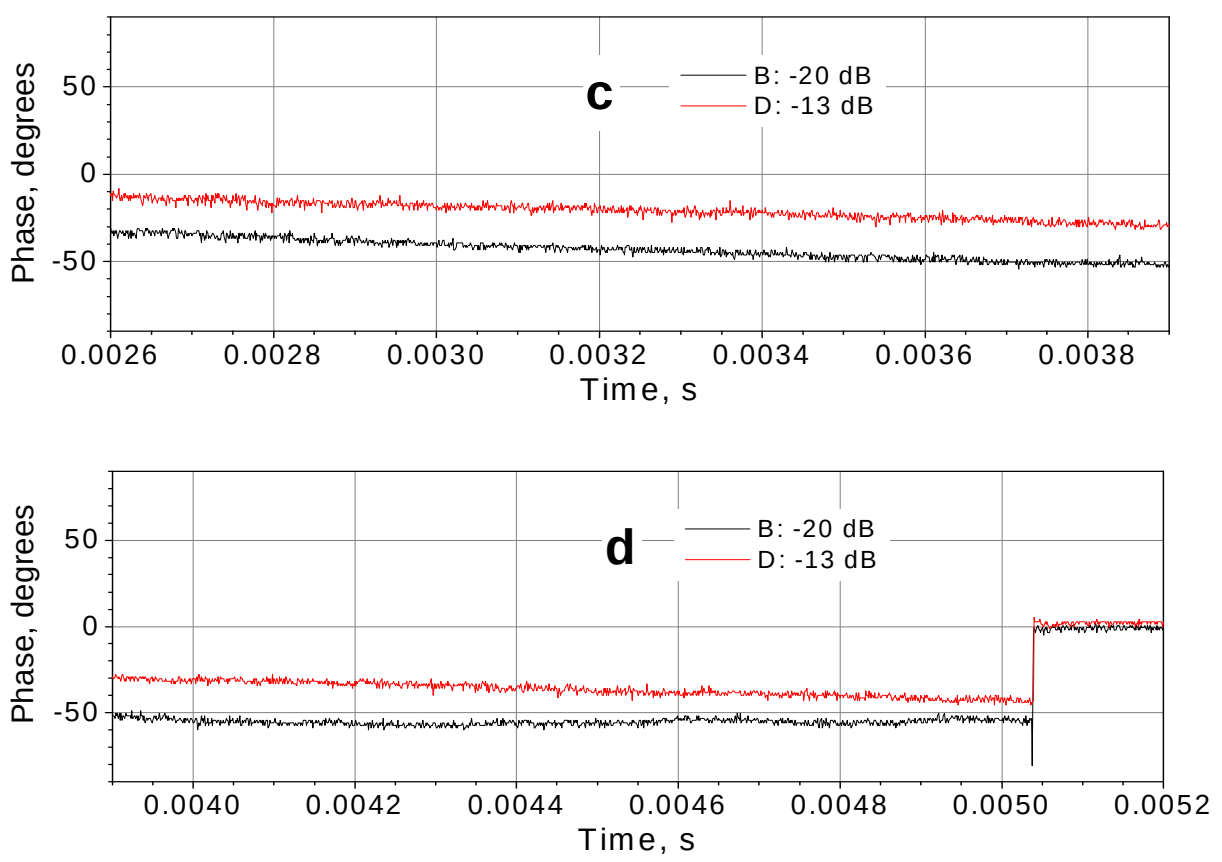

Figure 7, a-d: Measured phase deviations of the two-stage frequency-locked magnetron in pulsed regime. Traces B and D correspond to ratio $P_{Out-II}/P_{Lock-I} \approx$ 33.5 dB and 26.5 dB, respectively.

The plots show slow phase variations caused mainly by an intrapulse drop of the magnetron voltage. The phase noise value of the 2-stage magnetron is a few degrees. Time-to-lock of the 2-stage magnetron is $\leq$50 μs.

## CONCLUSION

We demonstrated the operation of the frequency locked CW single-stage and two-stage S-Band magnetrons in a pulsed regime. Both models provide smooth phase deviations caused in general by modulator parameters and the magnetron cathode properties. Instantaneous phase noise of the single stage and the two-stage magnetron are in range of few degrees. Measured ratios of the peak power to noise in the magnetron spectra are $\geq$45 dB. The ratio of the magnetron output power to locking power was $\approx$33.5 dB and $\approx$20 dB for the two-stage and single stage magnetrons respectively. These results were obtained in ordinary regimes at long pulse duration. The measured time-to-lock for single stage and two-stage frequency-locked magnetrons $\leq$50 μs is adequate for SC accelerators including the Project X pulsed linac. The obtained results show that the proposed concept of a two-stage (cascade) magnetron based on CW commercial tubes is a promising one for feeding SC intensity frontier accelerators.